\documentclass{article}
\usepackage{graphicx}
\usepackage{amssymb}
\usepackage{epstopdf}
\usepackage{amsmath}
\usepackage{natbib}
\usepackage{subfigure}
\usepackage{lscape, url}

\usepackage{pifont}
\usepackage{natbib}
\usepackage{latexsym}
\usepackage{yhmath}
\usepackage{bm}
\usepackage{epsfig}
\usepackage{fancyhdr}
\usepackage{sectsty,lscape,url}
\usepackage[hmargin=1in,vmargin=1in]{geometry}
\usepackage{sidecap}

\begin{document}

\title{Cliffs User Manual}
\author{Elena Tolkova \thanks{e.tolkova@gmail.com; elena@nwra.com} \\ NorthWest Research Associates, Bellevue, WA 98009-3027, USA}
\maketitle

\pagestyle{fancy}       
\lhead{}

\section{Overview}

Cliffs is an open-source relative of MOST (Method Of Splitting Tsunamis) numerical model. 
Cliffs solves the non-linear shallow-water equations in a domain with an arbitrary configuration of wet and dry nodes. In open water, Cliffs uses 1D finite-difference approximation VTCS-2 \cite[]{titov1995} modified as described in section Numerics in this paper, combined with the dimensional splitting \citep{strang, titov1998,titov97}. Cliffs solution on the land-water boundary is based on a staircase representation of topography and hence treating a moving shoreline as a moving vertical wall (a moving cliff), which gave the model its name \cite[]{cliffs}. 
The flow of the computations and input/output data types are similar to that in MOST version 4 (not documented; benchmarked for NTHMP in 2011 \cite[]{tolk-nthmp}), which was developed as an adaptation of a curvilinear version of the MOST model \cite[]{curvimost} to spherical coordinate systems arbitrary rotated on the Globe. For this Cliffs distribution, the computational flow of MOST-4 has been optimized to focus specifically on geophysical and Cartesian coordinate systems.
Cliffs performs computations in a single grid (further referred to as Master grid) given initial deformation of the free surface or the sea floor, and/or initial velocity field, and/or boundary forcing. It also computes boundary time-series input into any number of enclosed grids, to allow further refinement of the solution. Cliffs operates in either geographical (longitude, latitude) spherical or Cartesian coordinates, in 2D or 1D domains. 

All input and output data files, including bathymetry, should be in netcdf format. \\

Cliffs description is given in three parts. The first part, Running Options, presents a quick user manual. It explains command line options to run the executable, input parameters, and format of input/output data. 
The second part, Examples, describes sample modeling problems included with the code distribution. 
The last part, Numerics, lists numerical techniques utilized by Cliffs, and describes deviations from the earlier Cliffs version \cite[]{cliffs}. It concludes with discussing a specific numerical instability of the VTCS-2 algorithm introduced by abrupt changes in depth, and the ways to prevent the development of this instability.\\

Cliffs code was written in 03/2013-09/2014, and has been occasionally revisited thereafter. It is copyrighted under the terms of FreeBSD license. The author agrees to take full coding credit for the successful operation of Cliffs should it operate successfully. The author takes no responsibility for any type of actual or potential damages,  losses, or natural catastrophes caused by using the entire or any part of this code.

\section{Running Options}

Cliffs is coded in Fortran-95 and parallelized using OpenMP. Makefile included with the code distribution is used to compile the code on Mac with gfortran. Netcdf libraries should be installed.

\paragraph {The command line} to run the executable has five required and sixth optional parameter fields delimited by $< \dots >$ as follows:\\
./Cliffs $<$OuputDir/CaseTitle$>$ $ <$InputDataDir$>$  $<$BoundaryInputTitle or 0 if no boundary input$>$

 $<$InitialConditionsTitle or 0 if no initial conditions$>$  $<$ParameterDir/ParameterFile$>$ $<$short notes$>$\\
 Optional argument $<$short notes$>$ is a text string up to 200 characters in length (spaces are OK) to be printed at the top of a log file. 

\paragraph {The program output} is written in directory OutputDir. CaseTitle is assigned to all output files, which are: 
\begin{itemize}
\item
Netcdf binaries of velocities and elevation screenshots in Master grid saved to files named CaseTitle\_sea\_(u/v/h).nc, in the format specified below;
\item
Netcdf binaries of maximal water surface elevation and maximal current in Master grid in a file named CaseTitle\_maxwave.nc;
\item
Time histories of the free surface displacement $\eta$ and velocity components $u,v$ at virtual gages in a file named CaseTitle\_gages.nc;
\item
Netcdf binaries of boundary input time-series for the enclosed grids saved to files named CaseTitle\_BathyName\_(west/east/south/north).nc, one file for each boundary as implied by the name ending, four files per a grid named BathyName.
\item
Text log file documenting the program execution named CaseTitle\_log.txt. If $<$short notes$>$ in the command line are present, they are included at the top of the log file.
\end{itemize}

\begin{figure}[ht]
	\resizebox{1.1\textwidth}{!} 
			{\includegraphics{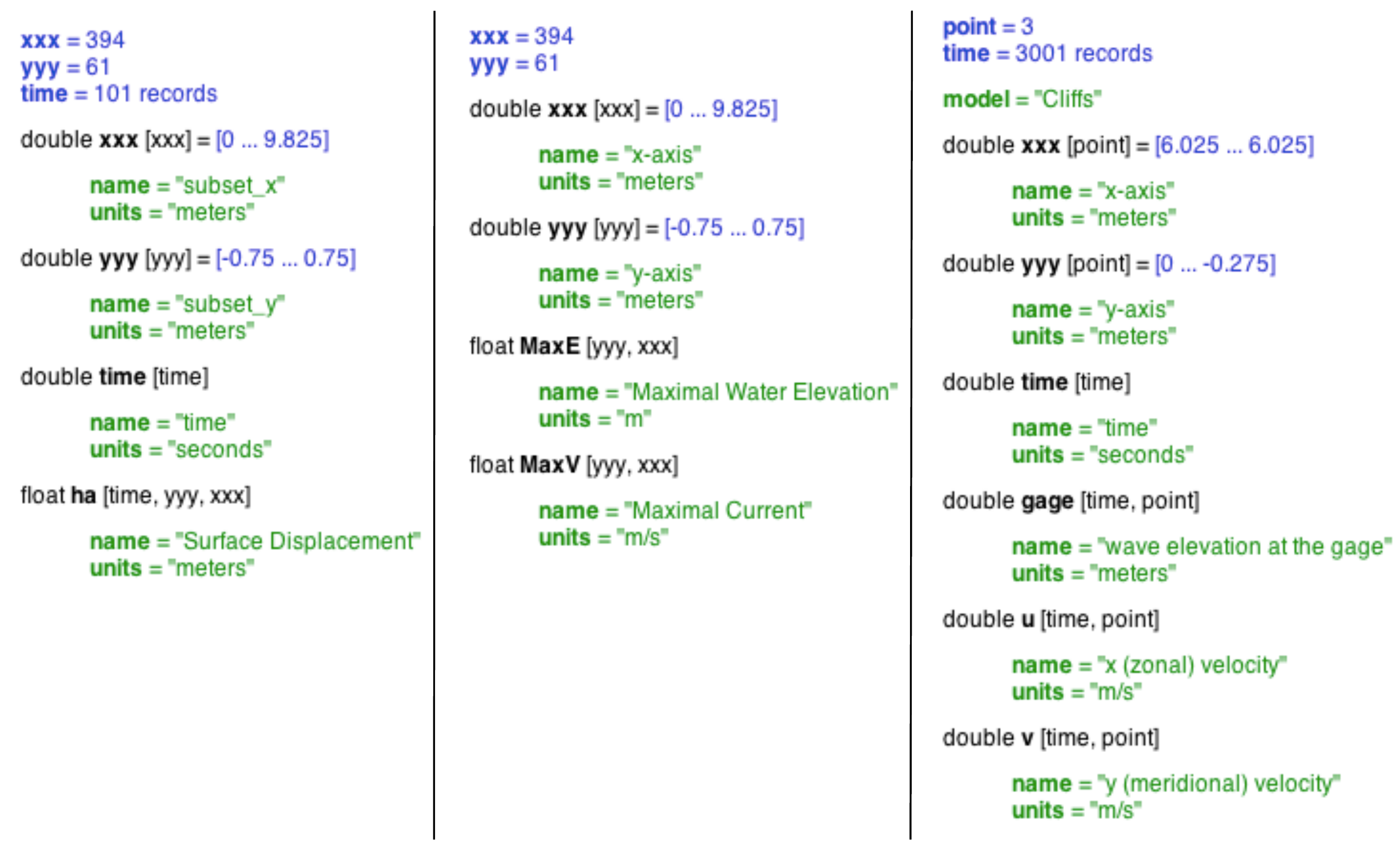}}
	\caption{Contents of sample Cliffs output as displayed by NetCDF QuickLook Plugin (http://tolkov.com/tvsoft/inetcdf/index.html):
	$name$\_sea.nc (left);  $name$\_maxwave.nc (middle);  $name$\_gages.nc (right).}
	\label{output}
\end{figure}

\paragraph{Netcdf structure} of Cliffs input/output and the naming convention is kept consistent with that of the MOST model, to facilitate Cliffs use by MOST users.
The screenshot output is defined against two spatial dimensions $lon$ or $xxx$ of length $nx$, $lat$ or $yyy$ of length $ny$, and one temporal dimension $time$ of length $nt$. During the computations, the $time$ dimension is defined as unlimited. The screenshot output is written in three netcdf files named CaseTitle\_sea\_(u/v/h).nc.  Each file contains a 3D float-type variable $ua(time,lat/yyy,lon/xxx)$, or $va(time,lat/yyy,lon/xxx)$, or $ha(time,lat/yyy,lon/xxx)$ of size $nt \times ny \times nx$, representing wave zonal (x-axis) current $u$, or meridional (y-axis) current $v$, or the free surface displacement $\eta$, respectively. Each screenshot file also contains three 1D double-type variables of the coordinates $lon/xxx$ and $lat/yyy$, and $time$, aligned with the same-name dimension (Fig.\ref{output}, left). 

Netcdf output of maximal water elevation and maximal current contains two vectors (type double) of spatial coordinates and two 2D data sets (type float) against the spatial dimensions $lon$ or $xxx$, and $lat$ or $yyy$ (Fig.\ref{output}, middle). 

Netcdf output of virtual gage records is written against two dimensions named $point$ (which contains a number of the observation point) and $time$. It contains two vectors of the gages coordinates (type double) along the dimension $point$, time vector (type double) along the dimension $time$, and three 2D data sets (type float) of the elevation and velocities $\eta, u, v$ recorded at each gage along dimensions $point$ and $time$ (Fig.\ref{output}, right). 

Each boundary time-series data file contains 3D double-type variable 
$vals(tim,uvq,pnt)$ of size $nt \times 3 \times np$ and 1D variable $time(tim)$ of length $nt$. Variable $vals$ (Netcdf id = 1) consists of $nt \times np$ trios of instant values of zonal (x) current, meridional (y) current, and surface elevation aligned with dimension $uvq$ of length 3, at time moments given by variable $time$ (Netcdf id = 2) aligned with dimension $tim$ of length $nt$, in the consequent nodes of the corresponding grid boundary aligned with  dimension $pnt$ of length $np$ (either $nx$ or $ny$).

Variables and dimensions in the screenshot data and initial conditions are sought by their names (case-sensitive); those in the boundary input are sought by their number  (Netcdf id).

\paragraph{Bathymetry data} on a structured rectangular grid are represented by a double or float vector of the longitude (or x-coordinate) values at the grid nodes, the same-type vector of the latitudes (or y-coordinates), and a double or float 2D variable of water depth or land elevation. The bathymetric data are sought by the variable Netcdf id, with the lon/x node coordinates being the first variable, lat/y being the second, and the bathy/topo data being the third.
Bathymetry data give the land elevation relative to the still sea surface, with the positive direction being down, that is, negative bathymetry values correspond to dry areas, while positive values provide an undisturbed water depth at each node. 
A 2D grid must have no less than 3 nodes in each direction. A 1D grid must still be defined against two netcdf dimensions, with the size of one dimension being equal 1. Nesting is not enabled for 1D grids, but varying spacing can be used instead.

\paragraph{The Units} in all data sets are $m$ for length, $s$ for time, and decimal degrees for longitude (positive toward East) and latitude (positive toward North).

\paragraph {The input to the computations} is sought in the directory InputDataDir. The input should be some of the following:
\begin{itemize}
\item
four (for a 2D domain) or two (for a 1D domain) boundary time-series files named \\
BoundaryIputTitle\_MasterGridName\_(west/east/south/north).nc, one file for each boundary (2D) or an edge node (1D); {\bf{and/or}} 
\item
initial deformation of the free surface or the sea floor \\
InitialConditionsTitle\_h.nc; {\bf{and/or}}
\item
either or both initial velocities of the water column\\
InitialConditionsTitle\_[u/v].nc.
\end{itemize}
The files providing initial velocities and surface elevation should have the same structure as the screenshot output.
In particular, the initial condition data set should contain time dimension. Initial conditions containing a single screenshot must have time dimension of length one.  
In the absence of the boundary input, the start time of the computations will be the time of the screenshot.  Initial conditions can also be read from the data set containing multiple screenshots (such as a data set created by the previous Cliffs run). This option is intended for the use in combination with the boundary input. The screenshot data set will be scanned for the frame taken at the time closest to the start of the boundary input. This frame will be used to initialize the wave state in the domain. The start time of the computations will be the start time of the boundary input. 

A grid on which the initial conditions are defined does not have to encompass the Master grid, or coincide with its nodes. The initial conditions will be interpolated onto a part of Master grid within the source grid, and set to zero outside it.

In the absence of the initial conditions, the computations are run under the boundary input into the still domain. If the boundary input is provided, then all four files must be present in a 2D case, and two files (east\&west, or south\&north) must be present in a 1D case. In a 1D case, the boundary input must still have the same structure as in a 2D case, with the length of the dimension $pnt$ equal 1, and the length of dimension $uvq$ equal 3, though only one velocity component will be used (which is $u$ if a 1-D domain occupies x-axis, and $v$ if it occupies y-axis). 

When computations are done in a nested grid configuration, the input is provided by the output from the previous simulation in the parent grid. Hence in this sequence of the simulations, $OutputDir$ and $InputDataDir$ refer to the same directory.   

\paragraph{The input Parameter File} is a text file with a list of computational parameters, which are:
\begin{enumerate}
\item
an integer indicating the type of a coordinate system: 1 - cartesian, otherwise - geophysical;
\item
name of Master grid bathymetry file
\item
$nests$ - number of grids enclosed in Master grid \footnote{Cliffs performs computations in a single grid (Master grid) and generates boundary input for each enclosed grid};
\item
names of bathymetry files for the enclosed grids, one name per line;
\item
$cuke$ - still sea threshold in the boundary input, to detect wave arrival on the boundary. The start time of the computations will be the moment when the surface displacement exceeded the still sea threshold. Set this parameter less or equal zero - to start the computations when the boundary record starts. This parameter is not used when the computations are initiated with initial conditions only.
\item
$ground$ - minimal flow depth. At this depth and below it, a node is considered dry; 
\item
$crough$ - friction coefficient (drag or Manning $n^2$, depending on an expression for the bottom friction with which the code is compiled);
\item
$itopo$ - an integer indicating a type of the land-water boundary: 0 - vertical wall at depth $dwall$, otherwise - land inundation enabled;
\item
$dwall$ - the depth to place a vertical wall; not used when $itopo \ne 0$;
\item
$dt$ - time increment for computations in Master grid;
\item
$steps\_total$ - total number of integration (time) steps to perform;
\item
$quake$ - an integer indicating whether an initial surface deformation should be applied to the sea floor ($quake=1$) or to the free surface (otherwise).
\item
an integer indicating whether the computations should stop if the boundary input stops (0), or continue for the requested number of time steps (otherwise); not used with the initial conditions input only, regardless of its value in the parameter file;
\item
$seaout$ - number of time steps between screenshots. Setting $seaout > steps\_total$ will suppress screenshot output, and the file trio CaseTitle\_sea\_* will not be generated;
\item
$lonsub$ - node number increment in $lon/x$ direction to sub-sample screenshots, resulting in the output at $1:lonsub:ma$ rows;
\item
$latsub$ - node number increment in $lat/y$ direction to sub-sample screenshots, resulting in the output at $1:latsub:na$ columns;
\item
$bndout$ -  number of time steps between saving wave variables along the boundaries of the enclosed grids; not used when no grids are enclosed;
\item
$maxout$ - number of time steps between updating the maximum water surface elevation. When $maxout>steps\_total$, the maximal surface height will be saved  only at the end of the computations; 
\item
$Ngages$ - number of virtual gages / observation points, to output the surface height time histories;
\item
$gout$ - number of time steps between outputs to the gage time histories;
\item
two-column list of the observation points indexes ($lon/x$ node number, $lat/y$ node number).
\end{enumerate}
Some parameters might not be used, but should still be present (except enclosed grids in field 4, or gage parameters in fields 20 and 21, if the respective number in field 3 or 19  is zero).

Bathymetry files (grids) must be in the directory ParameterDir. 

Sample parameter files are provided with the Examples.
\clearpage

\section{Examples}
More elaborated description of the next two canonical benchmark tests and their simulation with Cliffs can be found in \cite[]{cliffs}.
\subsection{Runup on a sloping beach}

\begin{figure}[ht]
	\resizebox{0.9\textwidth}{!} 
			{\includegraphics{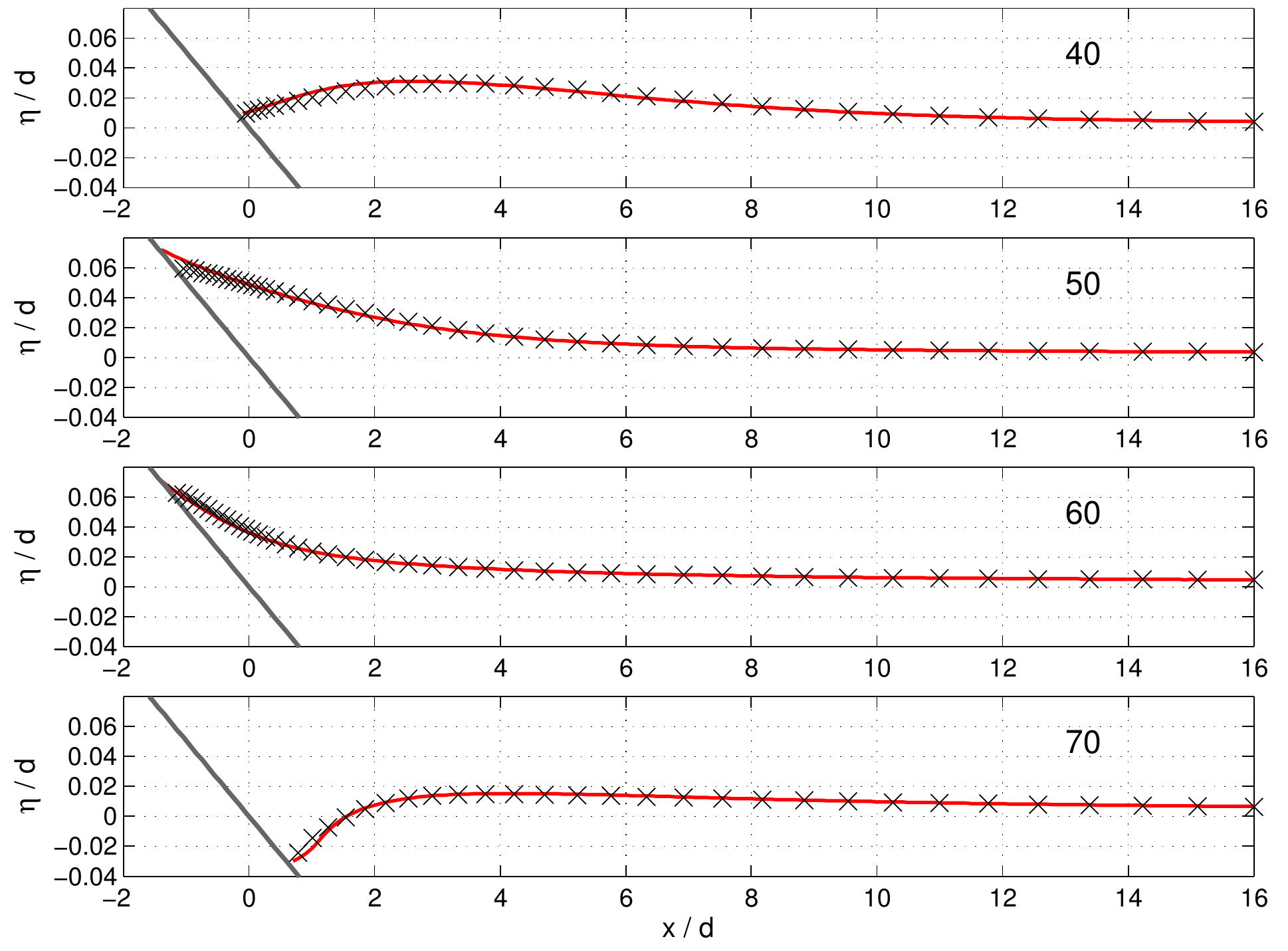}}
	\caption{Water surface profiles for an initial solitary wave $0.0185 d$ high climbing up a 1:19.85 beach at $t (g/d)^{1/2}=40,50,60,70$ (shown in a plot), black x - Cliffs numerical solution, red line - analytical solution. Plotted with Matlab script \emph{plotprofiles0185\_cliffs\_vs\_analyt.m} }
	\label{synolak}
\end{figure}


This example simulates a canonical problem of a solitary wave runup onto a plane beach with a 1:19.85 slope introduced by Synolakis (1987); and illustrates Cliffs operation in 1D configuration in Cartesian coordinates. The geometry of the beach, the lab experiment, and the wave-profile are described in many articles \citep{synolak1987, titov1995, synolak2007}. The cases with $H/d=0.0185$ and $H/d=0.3$, representing respectively a non-breaking and a severely breaking wave of initial height $H$ over depth $d$, are most commonly used for model verification \citep{liraich2002, nicolsky2011, nthmp} against laboratory and analytical data of Synolakis. Provided Cliffs set-ups to simulate these two cases include:
\begin{itemize}
\item
parameter file \emph{slpbeach\_params.txt}, shown in Table \ref{syn_params};
\item
bathymetry file \emph{slpbeach1D.nc}, with the grid spacing gradually varying from $d$ on the deep end to $0.1d$ near the beach, as described in \citep{tolk-nthmp,cliffs}; 
\item
initial conditions files \emph{solitary.0185\_h.nc} and \emph{solitary.0185\_u.nc}, to simulate a non-breaking wave case;
\item
initial conditions files \emph{solitary.30\_h.nc} and \emph{solitary.30\_u.nc}, to simulate a breaking wave case.
\end{itemize}

Should all input/output files and the executable be in the same directory, a command line to run the simulation can read:\\
\begin{tabular}{ccccccc}
./Cliffs &./sol0185& ./ &0 &solitary.0185& ./slpbeach\_params.txt & runup of a solitary wave 0.0185d high\\
OR & & & & & &\\
./Cliffs& ./sol30 &./& 0& solitary.30& ./slpbeach\_params.txt & runup of a solitary wave 0.3d high \\
\end{tabular}\\

Given the particular parameter file, Cliffs will generate four files (screenshots of velocity and elevation, maximal elevation, and a log file) with each run, for example:

\emph{sol0185\_sea\_h.nc}

\emph{sol0185\_sea\_u.nc}

\emph{sol0185\_maxwave.nc}

\emph{sol0185\_log.txt}\\ 

Phrase `runup ... high' in the command line is optional and, if present, will appear at the top of the log file.
Figure \ref{synolak} shows surface profiles extracted from Cliffs output \emph{sol0185\_sea\_h.nc} with a provided Matlab script (\emph{plotprofiles0185\_cliffs\_vs\_analyt.m}) atop the analytical solution (read from file \emph{canonical\_profiles.txt} provided by Dr. Dmitry Nicolsky, UAF for the  2011 NTHMP Model Benchmarking Workshop).

\begin{table}
\caption{Parameter file for simulating runup onto a sloping beach }
\begin{tabular}[b]{ll}
\hline
1&	1 - cartesian, otherwise-spherical\\
slpbeach1D.nc \\
0	&Number of grids enclosed in Master\\
0&	Still sea threshold on the boundary\\
0.003&	Minimal flow depth (m)\\
0&	friction coefficient (Manning n**2)\\
1&	topo flag: 0-wall, otherwise - land inundation\\
0.5&	Vertical wall, if any, at depth (m)\\
0.03&	time step (sec)   \\
2000&	total amount of steps\\ 
0&	1 - to deform bottom\\
1&	0 - to stop when boundary forcing stops\\
3&	steps between screenshots\\
1&	subsample screenshots in x\\
1&	subsample screenshots in y\\
1&	save feed into nested grids every n steps\\
500	&steps between maxwave updates\\
0&	N gages\\
\end{tabular}
\label{syn_params}
\end{table}

\clearpage

\subsection{Simulation of the 1993 Hokkaido-Nansei-Oki (Okushiri Island) tsunami}
\begin{figure}[ht]
	\resizebox{\textwidth}{!} 
		{\includegraphics{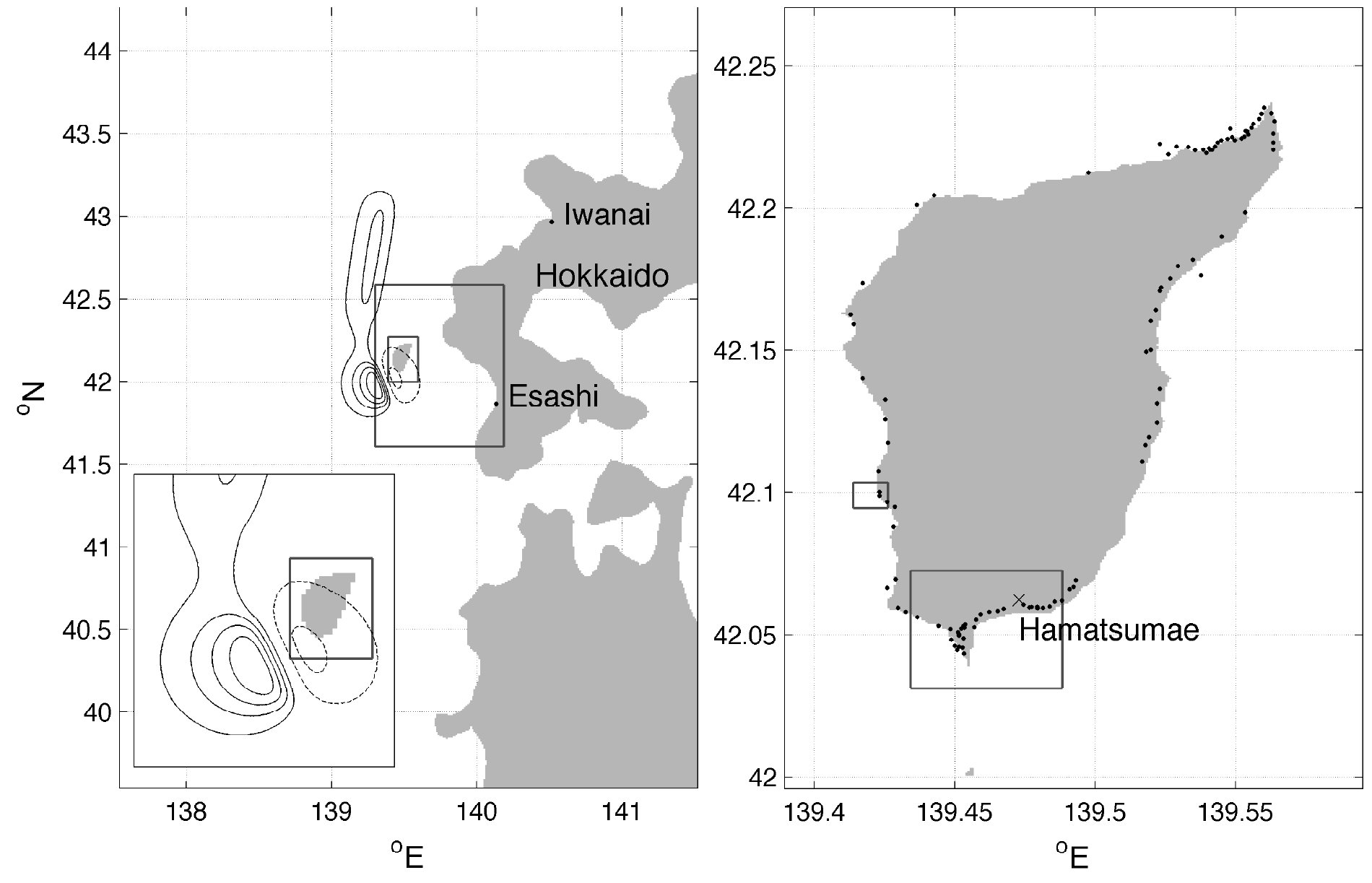}}
	\caption{Left: computational domain used to simulate the 1993 Hokkaido tsunami with contours of the two nested grids; initial sea surface deformation with contour lines at -1, -0.5, 1, 2, 3, 4 m levels, subsidence contours are shown with dashed lines, uplift with solid lines. The deformation area near Okushiri is zoomed-in in the bottom left corner. Right: 3-d nesting level grid around Okushiri island with contours of the 4-th level grids around Monai and Aonae; locations of field measurements.}
	\label{OKsetup}
\end{figure}

On July 12, 1993, the Mw 7.8 earthquake west of Okushiri island, Japan, generated a tsunami that has become a test case for tsunami modeling efforts \cite[]{takahashi}, \cite[]{synolak2007}, \cite[]{nthmp}. The complete set-up to simulate this event with Cliffs is provided. This example illustrates Cliffs operation in 2D configuration in geophysical coordinates with the use of nested grids, for simulating tsunami propagation from the source earthquake, and the consequent runup onto land. 

The event is simulated with the use of five grids at four levels of nesting, with resolution varying from 30 sec of Great arc (930 m) to 6 m. The outer grid (OK30s\_SSL2.1.nc) used to simulate the 1993 Okushiri tsunami and the initial sea surface deformation are shown in the left pane in Figure \ref{OKsetup}. The computations at a resolution of 30 sec of the Great arc were sequentially refined with nested grids spaced at 10 arc-sec (OK10s\_SSL2.1.nc, enclosed in OK30s\_SSL2.1.nc), 2 arc-sec around Okushiri island (OK02s\_SSL2.1.nc, enclosed in OK10s\_SSL2.1.nc), 15 m around Aonae peninsular (AO15m\_SSL2.1.nc, enclosed in OK02s\_SSL2.1.nc), and 6 m around Monai valley (MB06m\_SSL2.1.nc, enclosed in OK02s\_SSL2.1.nc as well). The grids contours are pictured in Figure \ref{OKsetup}.  
Computations in the first nesting level grid (OK30s\_SSL2.1.nc) are initiated with the initial deformation read from the file bottomdefBP9\_h.nc and applied to the sea floor. Computations in all other grids are run under the boundary input computed in the parent grid. The initial deformation of the sea floor is applied in each grid as well. 

To run the simulation, copy Cliffs executable to directory OkushiriTsunami. The directory also contains a subdir OkushiriGrids with all the grids and the corresponding parameter files, and a subdir Simulation with the bottom deformation file. The sequence of commands to perform the entire simulation (also contained in a script runcliffs2Okushiri) reads: \\

\begin{tabular}{llllll}
./Cliffs & Simulation/OK30 & Simulation/ & 0 &  bottomdefBP9 &  OkushiriGrids/paramsOK30s.txt \\
./Cliffs & Simulation/OK10 & Simulation/ & OK30 & bottomdefBP9 &  OkushiriGrids/paramsOK10s.txt \\
./Cliffs & Simulation/OK02 & Simulation/ & OK10 &  bottomdefBP9 &  OkushiriGrids/paramsOK02s.txt \\
./Cliffs & Simulation/AO  & Simulation/ & OK02 &  bottomdefBP9 &  OkushiriGrids/paramsAO15m.txt \\
./Cliffs & Simulation/MB & Simulation/ & OK02 &  bottomdefBP9 & OkushiriGrids/paramsMB06m.txt \\
\end{tabular}\\

Each run will result in populating directory Simulation with the results of the computations in a respective grid, and generated boundary input into the enclosed grid(s) if any.
As seen in the parameter files (tables 2-6), screenshot output interval $seaout$ is set greater than the total number of steps $steps\_total$, to suppress the screenshot output. Thus the computations in every grid result in saving the maximal wave height and in generating the boundary input into the next level grid. Computations in the first and second level grids also generate gage outputs with water level time histories near Iwanai and Esashi. The simulation time is 4 hrs, which is set by the simulation time in the first (30-sec) grid. The simulation time in other grids is set to longer according to the requested total number of time steps, but will be stopped earlier due to termination of the boundary input into that grid. A vertical wall is placed at 5 m depth in the 30-sec grid, and at 1 m depth in the 10-sec grid, while land inundation is permitted in the next three grids.

\begin{SCfigure}
\centering
	\caption{Maximal wave height in Aonae grid; colorbar - meters, black line - original shoreline before subsidence. Plotted with Matlab script \emph{aonae\_inundation.m}}
		\includegraphics[width=0.6\textwidth]{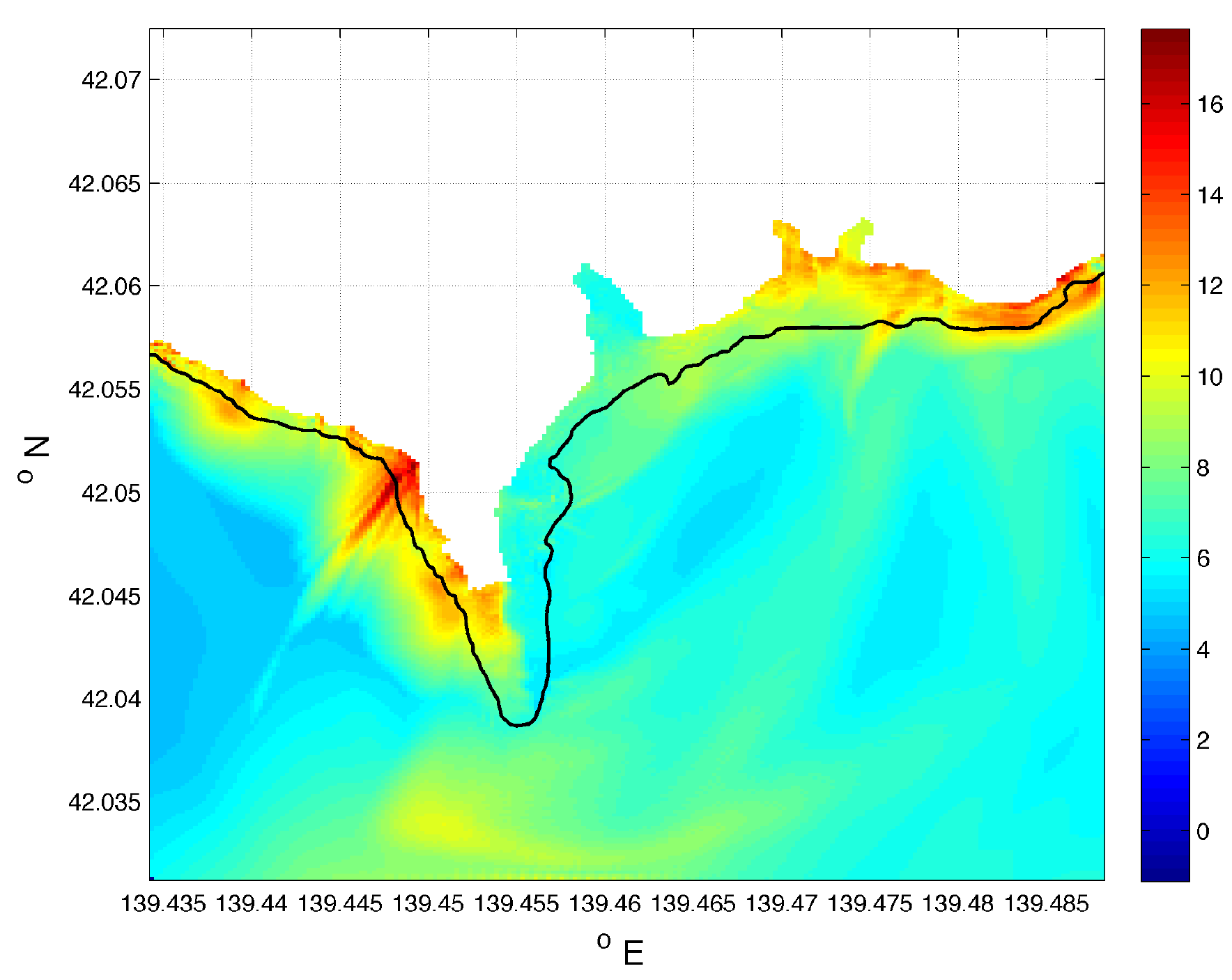}
	\label{aonae}
\end{SCfigure}

\begin{table}
\caption{Parameter file for simulating the Hokkaido tsunami, 1-st level grid }
\begin{tabular}[b]{ll}
\hline
0	&1 - cartesian, otherwise-spherical\\
OK30s\_SSL2.1.nc\\
1	&Number of grids enclosed in Master\\
OK10s\_SSL2.1.nc\\
0.001&	Still sea threshold on the boundary\\
1	&Minimal flow depth (m)\\
0.0009&	friction coefficient (Manning n**2)\\
0	&topo flag: 0-wall, otherwise - land inundation\\
5	&Vertical wall, if any, at depth (m)\\
4.0&	time step (sec)   \\
3600&	total amount of steps \\
1&	1 - to deform bottom\\
1&	0 - to stop when boundary forcing stops\\
5000	&steps between screenshots\\
1	&subsample screenshots in x\\
1	&subsample screenshots in y\\
1	&save feed into nested grids every n steps\\
500	&steps between maxwave updates\\
1	&N gages\\
5&  steps between outputs to gages\\
265 416& gage indexes\\
\end{tabular}
\label{ok1_params}
\end{table}

\begin{table}
\caption{Parameter file for simulating  the Hokkaido tsunami, 2-nd level grid }
\begin{tabular}[b]{ll}
\hline
0	&1 - cartesian, otherwise-spherical\\
OK10s\_SSL2.1.nc\\
1	&Number of grids enclosed in Master\\
OK02s\_SSL2.1.nc\\
0.001	&Still sea threshold on the boundary\\
0.5	&Minimal flow depth (m)\\
0.0009&	friction coefficient (Manning n**2)\\
0&	topo flag: 0-wall, otherwise - land inundation\\
1.0&	Vertical wall, if any, at depth (m)\\
1.5&	time step (sec)   \\
9600&	total amount of steps \\
1&	1 - to deform bottom\\
0&	0 - to stop when boundary forcing stops\\
10000&	steps between screenshots\\
1	&subsample screenshots in x\\
1	&subsample screenshots in y\\
1	&save feed into nested grids every n steps\\
500&	steps between maxwave updates\\
1	&N gages\\
12&   steps between outputs to gages\\
223 94& gage indexes\\
\end{tabular}
\label{ok2_params}
\end{table}

\begin{table}
\caption{Parameter file for simulating  the Hokkaido tsunami, 3-d level grid }
\begin{tabular}[b]{ll}
\hline
0	&1 - cartesian, otherwise-spherical\\
OK02s\_SSL2.1.nc\\
2	&Number of grids enclosed in Master\\
AO15m\_SSL2.1.nc\\
MB06m\_SSL2.1.nc\\
0.001	&Still sea threshold on the boundary\\
0.5	&Minimal flow depth (m)\\
0.0009&	friction coefficient (Manning n**2)\\
1	&topo flag: 0-wall, otherwise - land inundation\\
1	&Vertical wall, if any, at depth (m)\\
0.4	&time step (sec)   \\
36000&	total amount of steps \\
1&	1 - to deform bottom\\
0&	0 - to stop when boundary forcing stops\\
50000&	steps between screenshots\\
1	&subsample screenshots in x\\
1	&subsample screenshots in y\\
1	&save feed into nested grids every n steps\\
500	&steps between maxwave updates\\
0	&N gages\\
\end{tabular}
\label{ok3_params}
\end{table}

\begin{table}
\caption{Parameter file for simulating  the Hokkaido tsunami, 4-th level grid, Aonae }
\begin{tabular}[b]{ll}
\hline
0	&1 - cartesian, otherwise-spherical\\
AO15m\_SSL2.1.nc\\
0	&Number of grids enclosed in Master\\
0.001&	Still sea threshold on the boundary\\
0.1	&Minimal flow depth (m)\\
0.0009&	friction coefficient (Manning n**2)\\
1	&topo flag: 0-wall, otherwise - land inundation\\
1	&Vertical wall, if any, at depth (m)\\
0.5	&time step (sec)   \\
30000&	total amount of steps \\
1&	1 - to deform bottom\\
0&	0 - to stop when boundary forcing stops\\
50000	&steps between screenshots\\
1	&subsample screenshots in x\\
1	&subsample screenshots in y\\
1	&save feed into nested grids every n steps\\
500	&steps between maxwave updates\\
0	&N gages\\
\end{tabular}
\label{ok4_params}
\end{table}

\begin{table}
\caption{Parameter file for simulating the Hokkaido tsunami, 4-th level grid, Monai }
\begin{tabular}[b]{ll}
\hline
0	&1 - cartesian, otherwise-spherical\\
MB06m\_SSL2.1.nc\\
0	&Number of grids enclosed in Master\\
0.001	&Still sea threshold on the boundary\\
0.1	&Minimal flow depth (m)\\
0.0009	&friction coefficient (Manning n**2)\\
1	&topo flag: 0-wall, otherwise - land inundation\\
1	&Vertical wall, if any, at depth (m)\\
0.2	&time step (sec)   \\
50000	&total amount of steps \\
1&	1 - to deform bottom\\
0&	0 - to stop when boundary forcing stops\\
60000	&steps between screenshots\\
1	&subsample screenshots in x\\
1	&subsample screenshots in y\\
1	&save feed into nested grids every n steps\\
5000&	steps between maxwave updates\\
0	&N gages\\
\end{tabular}
\label{ok5_params}
\end{table}

\clearpage
\section{Numerics}

VTCS-2 numerical approximation was introduced by VT and CS in 1995 to model propagation and runup of 1D long waves under the framework of non-linear shallow-water theory \cite[]{titov1995}. VTCS-2 finite-difference scheme is described in detail in \citep{titov1995, titov1998}, and interpreted in a coding-friendly form in \cite[]{cliffs}. Burwell et al (2007) analyzed diffusive and dispersive properties of the VTCS-2 solutions. The present version of Cliffs uses a modification of the VTCS-2 scheme, described below.

A 1D algorithm can be efficiently applied to solving 2D shallow-water equations using well-known dimensional splitting method \citep{strang, yan, leveque2002}. 
In this way, the VTCS-2 scheme was extended to handle 2D problems in Cartesian coordinates \citep{titov1998}, geophysical spherical coordinates \cite[]{titov97} (the first mentioning of the MOST model), and in an arbitrary orthogonal curvilinear coordinates \cite[]{curvimost}. 

However, the dimensional splitting might result in underestimating an amplitude of a reflected wave in the MOST model, unless the reflecting boundary is aligned with either $x$ or $y$ coordinate axis. Slight modification to the reflective boundary conditions in MOST, equivalent to a half-node re-positioning of the reflecting wall within a cell, caused an appreciable difference in the results in some situations \cite[]{cliffs}. The modification was extended to include runup computations, by treating a moving shoreline as a moving vertical wall. Cliffs inundation algorithm is more compact and approximately 10\% more efficient computationally than the present MOST algorithm.

\subsection{Difference Scheme}
The shallow-water equations (SWE) solved in Cliffs are given below in matrix notation in the Cartesian coordinates:
\begin{equation}
W_t=A(W)W_x+B(W)W_y+C(W)
\label{eq1}
\end{equation}
where subscript denotes partial derivatives; $W=\begin{pmatrix} h & u & v \end{pmatrix}^T$ is a vector of state variables; $h$ is height of the water column; $u, \ v$ are particle velocity components in $x$ and $y$ directions;
\begin{equation}
A=-\begin{pmatrix} u & h & 0 \\ g & u & 0 \\ 0 & 0 & u \end{pmatrix}, \ B=-\begin{pmatrix} v & 0 & h \\ 0 & v & 0 \\ g & 0 & v \end{pmatrix}, \ C=\begin{pmatrix} 0 \\  gd_x-\alpha^x \\ gd_y-\alpha^y \end{pmatrix} ;
\label{eqABC}
\end{equation}
$g$ is acceleration due to gravity; $\begin{pmatrix} \alpha^x & \alpha^y \end{pmatrix}$ are the components of acceleration due to friction; $d$ is undisturbed water depth, or vertical coordinate of sea bottom measured down from the mean sea level (MSL). Negative values of $d$ correspond to dry land and give the land elevation relative to MSL. \\

Cliffs computes tsunami propagation using an efficient numerical method by Titov and Synolakis (1998). This method breaks the original SWE into separate problems of reduced complexity, to be solved sequentially. First, by making use of dimensional splitting, the original 2D problem becomes a sequence of 1D problems for the same state variables, to be solved row-wise and column-wise in an alternate manner: 
\begin{subequations}
\begin{equation}
W_t=A(W)W_x+C_1(W)
\label{eq2A}
\end{equation}
\begin{equation}
W_t=B(W)W_y+C_2(W)
\label{eq2B}
\end{equation}
\label{eq2AB}
\end{subequations}
where $C_1=\begin{pmatrix} 0 & gd_x-\alpha^x & 0 \end{pmatrix}^T$, $C_2=\begin{pmatrix} 0 & 0 & gd_y-\alpha^y \end{pmatrix}^T$.
Next, by transitioning to Riemann invariants 
\[
p=u+2 \sqrt{gh}, \ \ \ q=u-2 \sqrt{gh}
\]
with corresponding eigenvalues $ \lambda_{p,q}=u \pm \sqrt{gh}$, each 1D problem becomes three independent convection problems for a single variable each. The resulting problem set originating with a system \eqref{eq2A} follows:
\begin{subequations}
\begin{equation}
p_t =-(\lambda_1\cdot p_x - g d_x) -\alpha^x
\label{pq1}
\end{equation}
\begin{equation}
q_t=-( \lambda_2\cdot q_x - g d_x) -\alpha^x
\label{pq2}
\end{equation}
\begin{equation}
v_t = - u \cdot v_x
\label{pq3}
\end{equation}
\label{pq_most}
\end{subequations}

\paragraph{A derivation of a difference scheme} to integrate over one time step is given on an example of equation \eqref{pq1}. Using \eqref{pq1} with the Taylor expansion for $p(t)$ yields:
\begin{equation}
p^{n+1} =p^n-\Delta t \cdot ( \lambda p_x - g d_x)+\frac{{\Delta t}^2}{2} \cdot \lambda ( \lambda p_x - g d_x)_x   -\alpha \Delta t +O({\Delta t}^2)
\label{ptaylor}
\end{equation}
where $\Delta t$ is a time increment; all the right hand side variables are evaluated at a time step $n$. 
A non-trivial moment of constructing a finite-difference approximation for a term $\lambda p_x - g d_x$ is avoiding generation of artificial flow due to depth variations. The latter is achieved by a combination
\begin{equation}
Q(k,j)=\frac{1}{2} (\lambda_k+\lambda_j) \frac{p_k-p_j}{x_k-x_j}-g\frac{d_k-d_j}{x_k-x_j} 
\label{eqQ}
\end{equation}
In the steady state, $Q=0$, which ensures automatic preservation of the steady state. 

\paragraph{Two possible numerical approximations} for eq.\eqref{ptaylor} follow: 
\begin{equation}
p^{n+1}_j=p^n_j - \Delta t \cdot Q(j+1,j-1)+  \lambda_j  \Delta t^2 \cdot \frac{ \hat{Q}_j- \hat{Q}_{j-1}}{\Delta x_{j-1}+\Delta x_j}-\alpha_j{\Delta t}
\label{solver1}
\end{equation}
where $\Delta x_j=x_{j+1}-x_j$ is a space increment; $\hat{Q}_j=Q(j+1,j)$; and
\begin{equation}
p^{n+1}_j=p^n_j - \frac{\Delta t}{2} \cdot \left( \hat{Q}_{j-1} + \hat{Q}_j  \right)+  \lambda_j  \Delta t^2 \cdot \frac{ \hat{Q}_j- \hat{Q}_{j-1}}{\Delta x_{j-1}+\Delta x_j}-\alpha_j{\Delta t}
\label{solver2}
\end{equation}
The two stencils differ in that \eqref{solver1} uses a centered difference for the first derivatives at point $j$, whereas \eqref{solver2} uses an average between the left and right cells. Since term $Q$ is nonlinear with respect to depth and wave variables, 
\[
Q(j-1,j+1) \ne \frac{1}{2}  \left( \hat{Q}_{j-1} + \hat{Q}_j  \right) ,
\]
the two approximations yield noticeably different results for strongly nonlinear problems and/or over large depth variations. Scheme \eqref{solver1} is the original VTCS-2 scheme utilized in MOST and in the earlier Cliffs; scheme \eqref{solver2} is used in Cliffs presently.
Both solvers \eqref{solver1} and  \eqref{solver2} are first-order accurate in time and second-order accurate in space for a uniform grid spacing\footnote{Do $not$ attempt to ``improve" the accuracy of  \eqref{solver2} over varying grid by replacing an average 
$ \left( \hat{Q}_{j-1} + \hat{Q}_j  \right)/2$ 
with a weighted average 
$ \left( \Delta x_{j} \cdot \hat{Q}_{j-1} + \Delta x_{j-1} \cdot \hat{Q}_j  \right) / \left( \Delta x_{j-1} + \Delta x_j \right) $
 !}. 
In a basin with constant depth, each scheme is stable under the known limit on the Courant number: $|\lambda| \Delta t/ \Delta x \le 1$. 
\clearpage
\subsection{Inundation algorithm} 

\begin{figure}[ht]
	\resizebox{\textwidth}{!} 
			{\includegraphics{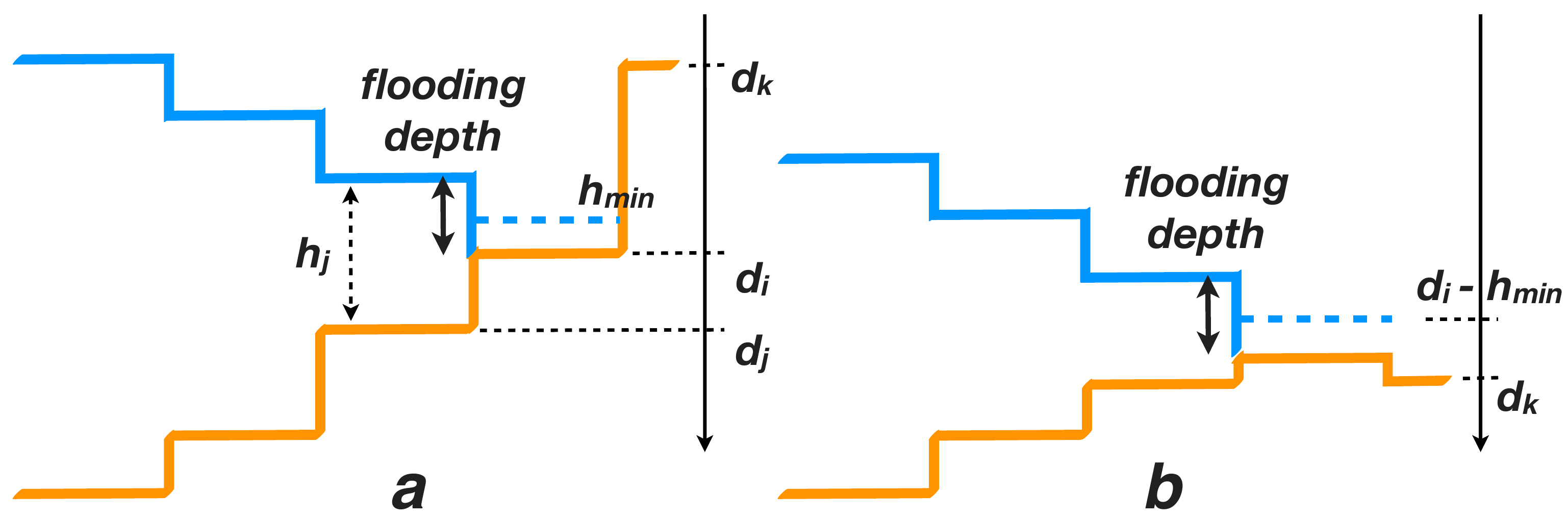}}
	\caption{From vertical wall to sloping beach.}
	\label{runup}
\end{figure}

Only cycle of the Cliffs solution procedure in a domain with wet and dry cells (discussed in detail in \cite[]{cliffs}, and slightly modified since then) comprises the next steps:
\renewcommand{\labelenumi}{(\alph{enumi})}
\begin{enumerate}
\item
The water height in each cell is compared with a threshold $h_{min}$. When the height is less then the threshold, the cell is marked as dry. 
\item
For each neighboring $i$-th dry and $j$-th wet cells ($j-i=\pm 1$), a decision is made on including the dry cell with the wet segment. If the flooding depth is greater than $h_{min}$, that is,
$h_j-d_j+d_i>h_{min}$, then $i$-th cell becomes wet with water column height $h_{min}$ and velocities equal to those in the adjacent wet cell (Figure \ref{runup}).
\item
The vertical interfaces between wet and dry cells are treated as vertical walls. 
\item
In all wet nodes, the solution for the next time step is computed according to \eqref{solver2}.
\item
The updated water height $h_i$ and velocity $u_i$ in a newly added wet cell on the shoreline are compared with the water height $h_j$ and velocity $u_j$ in the neighboring wet cell. Should the computation result in $h_i > h_j/2$, then the water height in $i$-th cell is set to $h_j/2$. Should the velocity $u_i$ exceed $u_j$, then it is assigned $u_i=u_j$. This empirical step prevents fragmentation and/or unrealistically rapid spreading of the runup cushion when the inundation occurs over very mild uphill slopes or downhill. It will be replaced in future releases.  
\end{enumerate}
Using reflecting boundary conditions on all wet segment boundaries yields an analogy of framing wet segments with moving mirrors (cliffs), which gave the model its name Cliffs. The run-up and run-down are made possible by expanding the wet area in step (b) and shrinking it in step (a). Since the difference scheme uses a centered 3-node pattern, the wave can propagate only as far as one cell per a time step, thus the wet area is also expanded by no more than one cell on each side. The runup occurs on an inserted cushion $h_{min}$ high, which is removed on rundown. \\

\clearpage
\subsection{Numerical stability}

In the presence of sharp changes in depth, the VTCS-2 solutions sometimes develop a specific, slowly growing instability. This instability is not described in the literature, and therefore it is discussed here. 
Let us compute one-time-step evolution of an infinitesimally small initial pulse with height (surface displacement) $\eta_j$ and velocity $u_j$ applied in a single node $j$ in a still basin. For simplicity, assume uniform spacing $\Delta x$. Then
\[
\hat{Q}_{j-1}=\frac{1}{\Delta x} \left( \frac{1}{2}u_j \left( 3 \sqrt{gd_j}- \sqrt{gd_{j-1}} \right) + g \eta_j \right)+ O^2,
\]
\[
\hat{Q}_j=-\frac{1}{\Delta x} \left( \frac{1}{2}u_j \left( 3 \sqrt{gd_j}- \sqrt{gd_{j+1}} \right) +g \eta_j \right) + O^2,
\]
where $O^2$ refers to the second order quantities in $u$ and $\eta$,
\[
\frac{1}{2}(\hat{Q}_{j-1}+\hat{Q}_j)=\frac{u_j}{4 \Delta x} \left( \sqrt{gd_{j+1}}- \sqrt{gd_{j-1}} \right) + O^2,
\]
whereas
\[
Q(j+1,j-1)=0.
\]
Substituting the above into \eqref{solver1} yields
\begin{equation}
p_j^{+1}=p_j -\sqrt{gd_j} {\frac{ {\Delta t}^2}{{\Delta x}^2}}\left[ \frac{u_j}{4} \left( 6 \sqrt{gd_j}- \sqrt{gd_{j-1}}- \sqrt{gd_{j+1}} \right)+g\eta_j \right] +O^2.
\label{p1}
\end{equation}
Substituting into \eqref{solver2} yields
\begin{equation}
p_j^{+1}=p_j -\sqrt{gd_j} {\frac{ {\Delta t}^2}{{\Delta x}^2}}\left[ \frac{u_j}{4} \left( 6 \sqrt{gd_j}- \sqrt{gd_{j-1}}- \sqrt{gd_{j+1}} \right)+g\eta_j \right] 
-\frac{u_j \Delta t}{4\Delta x} (\sqrt{gd_{j+1}}-\sqrt{gd_{j-1}})+O^2.
\label{p2}
\end{equation}
Solution for the corresponding $q_j^{+1}$ can be obtained from \eqref{p1}/ \eqref{p2} by reverting signs of radicals. 
Once values $p$ and $q$ have been updated, the corresponding state variables are recovered as 
\begin{subequations}
\begin{equation}
u=(p +q)/2
\label{stateu}
\end{equation}
\begin{equation}
h=(p-q)^2/ 16g
\label{stateh}
\end{equation}
\label{stateuh}
\end{subequations}
Then according to \eqref{solver1}:
\begin{subequations}
\begin{equation}
u_j^{+1}  = \left( 1- \beta^2_j f_j \right) u_j+ O^2
\label{eq:u}
\end{equation}
\begin{equation}
\eta_j^{+1}  =  \left( 1- \beta^2_j \right) \eta_j+ O^2
\label{eq:et}
\end{equation}
\label{eq:uet}
\end{subequations}
According to \eqref{solver2}:
\begin{subequations}
\begin{equation}
u_j^{+1}  = \left( 1- \beta^2_j f_j \right) u_j+ O^2
\label{eq:u2}
\end{equation}
\begin{equation}
\eta_j^{+1}  =  \left( 1- \beta^2_j \right) \eta_j - \frac{\beta_j}{4 \sqrt{g}} (\sqrt{d_{j+1}}-\sqrt{d_{j-1}}) u_j + O^2
\label{eq:et2}
\end{equation}
\label{eq:uet2}
\end{subequations}
where $\beta_j=\Delta t {\sqrt{gd_j}} / \Delta x $ is the Courant number associated with $j$-th node; factor $f_j$ is equal to
\begin{equation}
f_j=\frac {6 \sqrt{d_j} - \sqrt{d_{j-1}} - \sqrt{d_{j+1}}} {4 \sqrt{d_j}}
\label{eq:f}
\end{equation}
First, we observe the difference between the solutions by \eqref{solver1} and \eqref{solver2} in the presence of depth variations. Next, 
as seen from (\ref{eq:u})/(\ref{eq:u2}), either solution over varying bottom is likely to become unstable whenever factor $f_j$ is negative. 
Thus a condition $f>0$ restricting depth variations as
\begin{equation}
6 \sqrt{d_j} > \sqrt{d_{j-1}} + \sqrt{d_{j+1}}
\label{fcond}
\end{equation}
is necessary to ensure stability of either solver, in addition to CFL condition. \\

Being limited to a very particular case, \eqref{fcond} does not provide a sufficient stability condition. To strengthen the condition, one might want to restrict  depth variations on the left and the right of $j$-th node independently. Let us expand factor $f_j$ as
\begin{equation}
f_j=\frac{1}{2} (1+r_l) +\frac{1}{2}(1+r_r)
\label{f_expnd}
\end{equation}
where
\begin{equation}
r_l=\frac{\sqrt{d_j}-\sqrt{d_{j-1}}}{2\sqrt{d_j}} , \ \ r_r=\frac{\sqrt{d_j}-\sqrt{d_{i+j}}}{2\sqrt{d_j}} 
\label{rr}
\end{equation}
An empirical stability condition which ensures positive $f$ follows:
\begin{equation}
1+r_r \ge \epsilon>0 \ \ \ AND \ \ \ 1+r_l \ge \epsilon>0,
\label{rstab}
\end{equation}
or
\begin{equation}
\max \{ d_{i-1}/d_i, \ \ d_{i+1}/d_i \} \le \alpha^2,
\label{dstab}
\end{equation}
where $\alpha=3-2\epsilon$, $0<\epsilon<1$, $1<\alpha <3$. Typically, $\alpha=2$ ($\epsilon=0.5$) is sufficient. $\alpha=1$ can only be met in a basin with constant depth.

\paragraph{It is recommended to pre-process a digital map of a sea floor} to impose restrictions \eqref{rstab}/\eqref{dstab} both row-wise and column-wise, as follows:
given $\alpha$, at any node $j$ where (\ref{dstab}) does not hold, the depth value is to be replaced with 
\begin{equation}
d_j = (\sqrt{d_{j-1}}+\sqrt{d_{j+1}})^2/4
\label{newdw}
\end{equation}
in an open water, and with
\begin{equation}
d_j = d_w/\alpha^2
\label{newdc}
\end{equation}
next to the coastline, where $d_w$ is the depth at the wet neighbor of the $j$-th node.
Substituting updated depth \eqref{newdw} into \eqref{rr}, and the later ones into condition \eqref{rstab} results in
\begin{equation}
(1-\epsilon)\sqrt{d_{i-1}} + (3-\epsilon)\sqrt{d_{i+1}} > 0, \ \ \ (1-\epsilon)\sqrt{d_{i+1}} + (3-\epsilon)\sqrt{d_{i-1}} > 0
\end{equation}
which satisfies \eqref{rstab} for any $\epsilon<1$. \\

\paragraph{Program \emph{depth\_ssl}} (supplied with Cliffs distribution) can be used to perform the suggested pre-processing of a depth map. A command line to run the program should read:\\

./depth\_ssl \ \ $<$InputDir/InputBathyName.nc $>$ \ \  $<$OutputDir/$>$ \ \  $< n >$ \ \  $[<$OutputBathyName.nc$>]$\\

The third argument is an integer indicating that the program should inquire about the desired limit $\alpha$ (between 1.5 and 2.5), and the minimal water depth (m) in wet nodes. If this argument is set to 0, the default $\alpha=2$ and $min\_depth=0.1$ m will be used.
If the last argument is omitted, the output bathy file will be created, named InputBathyName\_ssl.nc, in the directory OutputDir. 

Bathymetry files should/will be in the bathy/topo netcdf format described in section 2.

\end{document}